\documentclass[prd,amsmath,amssymb,showpacs,superscriptaddress,nofootinbib,twocolumn]{revtex4-1}
\usepackage{graphicx}
\usepackage{epsfig,graphics,subfigure,psfrag,amsmath,amssymb}
\usepackage{dcolumn}
\usepackage{bm}
\usepackage{overpic}
\usepackage[english]{babel}
\usepackage{color}
\usepackage{multirow}
\usepackage{colortbl}
\usepackage{epstopdf}
\usepackage[colorlinks,linkcolor=blue,anchorcolor=blue,citecolor=blue]{hyperref}
\usepackage{gensymb}
\usepackage{ulem}

\def \stat{\mbox{$\,$(stat.)}}
\def \syst{\mbox{$\,$(syst.)}}

\begin{document}

\title{\boldmath Observation of the decays $\chi_{cJ}\to\Sigma^{0}\bar{p}K^{+}+{\rm c.c.}~(J=0, 1, 2)$}

\author{
\begin{small}
\begin{center}
M.~Ablikim$^{1}$, M.~N.~Achasov$^{10,c}$, P.~Adlarson$^{67}$, S. ~Ahmed$^{15}$, M.~Albrecht$^{4}$, R.~Aliberti$^{28}$, A.~Amoroso$^{66A,66C}$, M.~R.~An$^{33}$, Q.~An$^{63,50}$, X.~H.~Bai$^{57}$
% ...
\\
\vspace{0.2cm}
(BESIII Collaboration)\\
\vspace{0.2cm} {\it
$^{1}$ Institute of High Energy Physics, Beijing 100049, People's Republic of China\\
% ...
$^{4}$ Bochum Ruhr-University, D-44780 Bochum, Germany\\
% ...
$^{10}$ G.I. Budker Institute of Nuclear Physics SB RAS (BINP), Novosibirsk 630090, Russia\\
% ...
$^{15}$ Helmholtz Institute Mainz, Johann-Joachim-Becher-Weg 45, D-55099 Mainz, Germany\\
% ...
$^{28}$ Johannes Gutenberg University of Mainz, Johann-Joachim-Becher-Weg 45, D-55099 Mainz, Germany\\
% ...
$^{33}$ Liaoning Normal University, Dalian 116029, People's Republic of China\\
% ...
$^{50}$ State Key Laboratory of Particle Detection and Electronics, Beijing 100049, Hefei 230026, People's Republic of China\\
% ...
$^{57}$ University of Jinan, Jinan 250022, People's Republic of China\\
% ...
$^{63}$ University of Science and Technology of China, Hefei 230026, People's Republic of China\\
% ...
$^{66}$ University of Turin and INFN, INFN, I-10125, Turin, Italy\\
$^{67}$ Uppsala University, Box 516, SE-75120 Uppsala, Sweden\\
% ...
$^{c}$ Also at the Novosibirsk State University, Novosibirsk, 630090, Russia\\
% ...
}
\end{center}
\vspace{0.4cm}
\end{small}
}

\begin{abstract}
  The decays $\chi_{cJ}\to\Sigma^{0}\bar{p}K^{+}+{\rm c.c.}~(J = 0, 1, 2)$ are studied via the radiative transition $\psi(3686)\to\gamma\chi_{cJ}$ based on a data sample of $(448.1 \pm 2.9)\times10^{6}$ $\psi(3686)$ events collected with the BESIII detector. The branching fractions of $\chi_{cJ}\to\Sigma^{0}\bar{p}K^{+}+{\rm c.c.}~(J = 0, 1, 2)$ are measured to be $(3.03 \pm 0.12\pm 0.15)\times10^{-4}$, $(1.46 \pm 0.07\pm 0.07)\times10^{-4}$, and $(0.91 \pm 0.06\pm 0.05)\times10^{-4}$, respectively, where the first uncertainties are statistical and the second are systematic. In addition, no evident structure is found for excited baryon resonances on the two-body subsystems with the limited statistics.

\end{abstract}

\maketitle

\section{\boldmath Introduction}

 The P-wave charmonia $\chi_{cJ}~(J=0, 1, 2)$ have been observed experimentally for a long time, however, most decay modes of them are still unknown. Though $\chi_{cJ}$ can not be directly produced via electron-positron annihilation into a virtual photon, radiative decays of the $\psi(3686)$ into $\chi_{cJ}$ states make up about $10\%$ of the total decay width of the $\psi(3686)$ for each $\chi_{cJ}$~\cite{PDG}. Thus, the large $\psi(3686)$ data sample containing $(448.1 \pm 2.9)\times10^{6}$ events at BESIII can ideally be used to investigate $\chi_{cJ}$ decays~\cite{Asner2009,Ablikim2019}.

 Many two-body decays of $\chi_{cJ}\to B\bar{B}$ have been observed in experiments, but three-body decays of $\chi_{cJ}\to B\bar{B}M$ are much less measured ($B$ stands for a baryon, $M$ stands for a meson), while the latter have advantages to search for and study excited baryons due to larger freedom of quantum numbers. For example, some experiments reported two excited $\Sigma$ resonances around 1670~MeV/$c^{2}$, which have the same mass and $J^{PC}$ quantum numbers but very different decay products and angular distributions~\cite{sigmapi,sigma1,sigma2,sigma4}. Further experimental information will shed light on the understanding of these states.

 The decays of $\chi_{cJ}\to\Sigma^{+}\bar{p}K^{0}_{S}+{\rm c.c.}~(J = 0, 1, 2)$ have been measured at BESIII~\cite{isos}, which implies the existence of isospin conjugate channels $\chi_{cJ}\to\Sigma^{0}\bar{p}K^{+}+{\rm c.c.}~(J=0, 1, 2)$. The decays of $\chi_{cJ}\to\Sigma^{0}\bar{p}K^{+}+{\rm c.c.}~(J=0, 1, 2)$ can be used to search for the excited $\Sigma$ resonances and understand their properties.

 In this analysis, we present a study of $\psi(3686)\to\gamma\chi_{cJ}$, $\chi_{cJ}\to\Sigma^{0}\bar{p}K^{+}+{\rm c.c.}~(J = 0, 1, 2)$, where $\Sigma^{0}$ is reconstructed in its dominant decay mode $\Sigma^{0}\to\gamma\Lambda$ with $\Lambda\to p\pi^{-}$. Throughout the analysis, unless otherwise noted, charge-conjugation is implied.

\section{\boldmath BESIII Detector}

 The BESIII detector~\cite{BES3} records symmetric $e^+e^-$ collisions provided by the BEPCII storage ring~\cite{Yu:IPAC2016-TUYA01}, which operates with a peak luminosity of $1\times10^{33}$~cm$^{-2}$s$^{-1}$ in the center-of-mass energy range from 2.0 to 4.7~GeV.
 The cylindrical core of the BESIII detector consists of a helium-based multilayer drift chamber (MDC), a plastic scintillator time-of-flight system (TOF), and a CsI (Tl) electromagnetic calorimeter (EMC), which are all enclosed in a superconducting solenoidal magnet providing a 1.0~T  magnetic field.  The solenoid is supported by an octagonal flux-return yoke with resistive plate counter muon identifier modules interleaved with steel. The charged-particle momentum resolution at 1~GeV/$c$ is 0.5\%, and the $dE/dx$ resolution is 6\% for the electrons of 1~GeV/$c$ momentum. The EMC measures photon energies with a resolution of 2.5\% (5\%) at 1~GeV in the barrel (end-cap) region. The time resolution of the TOF barrel part is 68~ps, while that of  the end-cap part is 110~ps.

\section{\boldmath Data set and Monte Carlo Simulation}

 This analysis is based on a sample of $(448.1\pm2.9)\times10^{6}$ $\psi(3686)$ events~\cite{psidata} collected with the BESIII detector.

 Simulated data samples produced with a {\sc geant4}-based~\cite{geant4} Monte Carlo (MC) package, which includes the geometric description of the BESIII detector and the detector response, are used to determine detection efficiencies and to estimate backgrounds. The simulation models the beam energy spread and initial state radiation (ISR) in the $e^+e^-$ annihilations with the generator {\sc kkmc}~\cite{kkmc,kkmc2}. The inclusive MC sample includes $506\times 10^6$ $\psi(3686)$ events, the ISR production of the $J/\psi$, and the continuum processes incorporated in {\sc kkmc}. The known decay modes are modelled with {\sc evtgen}~\cite{evtgen,evtgen2} using branching fractions taken from the Particle Data Group~\cite{PDG}, and the remaining unknown charmonium decays are modelled with {\sc lundcharm}~\cite{lundcharm}. Final state radiation~(FSR) from charged particles is incorporated using the {\sc photos} package~\cite{photos}.

 The decays of $\psi(3686)\to \gamma\chi_{cJ}(J=0, 1, 2)$ are simulated following Ref.~\cite{M2}, in which the magnetic-quadrupole (M2) transition for $\psi(3686)\to\gamma\chi_{c1,2}$ and the electric-octupole (E3) transition for $\psi(3686)\to\gamma\chi_{c2}$ are considered in addition to the dominant electric-dipole (E1) transition. The three-body decays $\chi_{cJ}\to \Sigma^{0}\bar{p}K^{+}+$~c.c. are generated evenly distributed in phase-space (PHSP).

\section{\boldmath Event selection and background analysis}

 For $\psi(3686)\to\gamma\chi_{cJ}$, $\chi_{cJ}\to \Sigma^{0}\bar{p}K^{+}$ with $\Sigma^{0}\to \gamma\Lambda$ and $\Lambda\to p\pi^{-}$, the final state consists of $p\bar{p}K^+\pi^-\gamma\gamma$. Charged tracks must be in the active region of the MDC, corresponding to $|\cos\theta|<$ 0.93, where $\theta$ is the polar angle of the charged track with respect to the symmetry axis of the detector. For the two charged tracks from the $\Lambda$ decay, the distance between their point of closest approach and the primary vertex is required to be less than 20~cm along the beam direction, and less than 10~cm in the plane perpendicular to the beam direction. For the remaining charged tracks, the same distance is required to be less than 10~cm along the beam direction and less than 1~cm in the plane perpendicular to the beam direction. The total number of charged tracks needs to be equal to or greater than four.

 The TOF and $dE/dx$ information is used to calculate a particle identification (PID) likelihood ($P$) for the hypotheses that a track is a pion, kaon, or proton. Tracks from the primary vertex are required to be identified as either an anti-proton ($P(p)>P(K)$ and $P(p)>P(\pi)$) or a kaon ($P(K)>P(p)$ and $P(K)>P(\pi)$). In case of daughter particles of a $\Lambda$ candidate, the track with the larger momentum is identified as the proton, and the other is identified as the pion. For each candidate event, exactly one $\bar{p}, K^{+}$, and $p$, $\pi^-$ from the $\Lambda$ decay are required.

\begin{figure}[htbp]
  \centering
  \subfigure{\includegraphics[width=7.5cm,height=5.0cm]{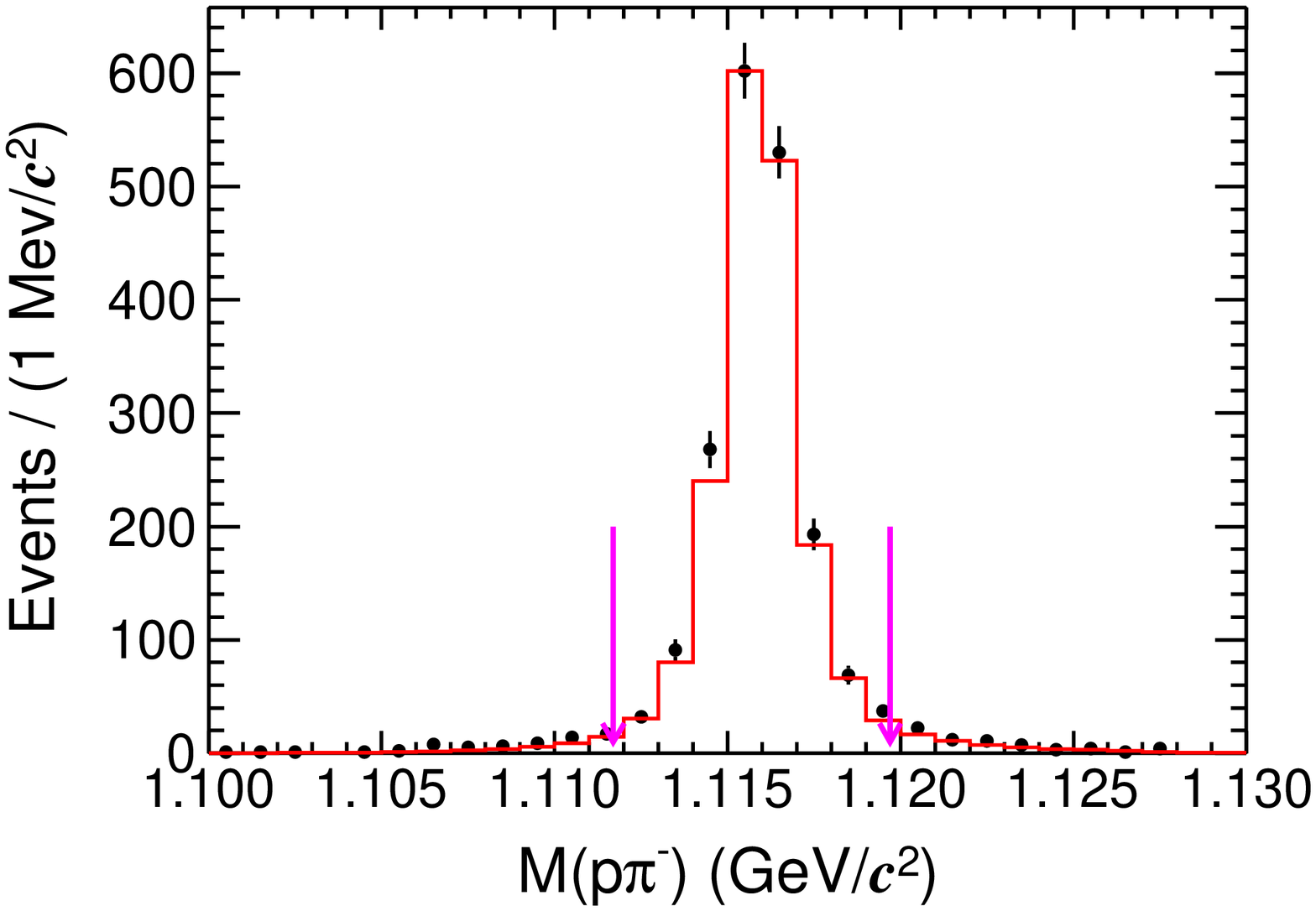}\put(-40,110){(a)} }
  \vspace{-0.5cm}
  \subfigure{\includegraphics[width=7.5cm,height=5.0cm]{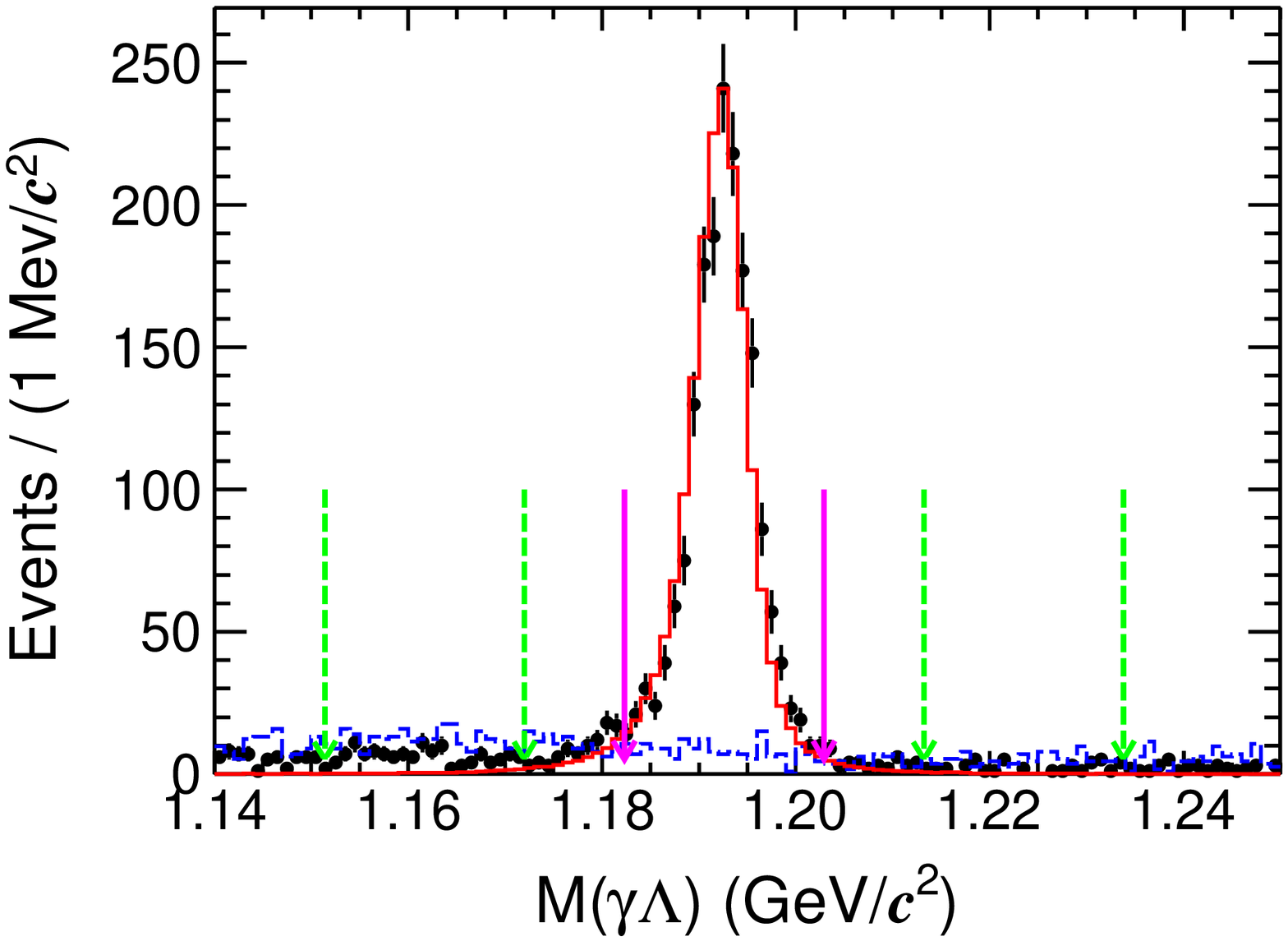}\put(-40,110){(b)} }
  {\caption{(a) The distribution of the $p\pi$ invariant mass. (b) The distribution of the $\gamma\Lambda$ invariant mass. The solid arrows respectively show the $\Lambda$ and $\Sigma^{0}$ mass windows, and the dashed arrows show the $\Sigma^{0}$ sideband mass regions. Dots with error bars are data, the histograms with solid lines represent signal MC simulations, and the dashed line in (b) is the background contribution from the inclusive MC sample scaled to the total number of $\psi(3686)$ events.}
  \label{LambaSigma}}
\end{figure}

 For all combinations of positively and negatively charged tracks, secondary vertex fits are performed~\cite{senvtx}, and the combination with the smallest $\chi^{2}_{\Lambda}$ is retained as the $\Lambda$ candidate. In addition, the ratio of the decay length ($L$) to its resolution ($\sigma_L$) is required to be larger than zero. The mass distribution of the reconstructed $\Lambda$ candidates is shown in Fig.~\ref{LambaSigma}(a). A mass window of $|M_{p\pi^{-}}-m_{\Lambda}|< 0.004$ GeV/$c^{2}$ is required to select the $\Lambda$ signal events, where $M_{p\pi^{-}}$ is the invariant mass of selected proton-pion pairs and $m_{\Lambda}$ is the nominal mass of $\Lambda$ taken from the PDG~\cite{PDG}.

 Photon candidates are reconstructed from the energy deposition in the EMC crystals produced by electromagnetic showers. The minimum energy requirement for a photon candidate is 25~MeV in the barrel region ($|\cos\theta|<0.80$) and 50~MeV in the end-cap region ($0.86<|\cos\theta|<0.92$). To eliminate showers originating from charged particles, a photon cluster must be separated by at least $10^\circ$ from any charged tracks. The time-information of the shower is required to be within 700~ns from the reconstructed event start-time to suppress noise and energy deposits unrelated to the event. The total number of photons is required to be at least two. To reduce background events from $\pi^{0}\rightarrow\gamma\gamma$, we require $|M_{\gamma\gamma}-m_{\pi^{0}}|>0.015$~GeV/$c^{2}$.

 A four-constraint (4C) kinematic fit imposing four-momentum conservation is performed using the $p\bar{p}K^+\pi^-\gamma\gamma$ hypothesis. If there are more than two photon candidates in one event, the combination with the smallest $\chi^{2}_{\rm 4C}$ is retained, and its $\chi^{2}_{\rm 4C}$ is required to be smaller than those for the alternative $p\bar{p}K^{+}\pi^{-}\gamma$ and $p\bar{p}K^{+}\pi^{-}\gamma\gamma\gamma$ hypotheses. In addition, the value of $\chi^{2}_{\rm 4C}$ is required to be less than 40. For the selected signal candidates, the $\gamma\Lambda$ combination with the invariant mass closest to the nominal $\Sigma^{0}$ mass according to the PDG~\cite{PDG} is taken as the $\Sigma^{0}$ candidate. The distribution of the $\gamma\Lambda$ invariant mass is shown in Fig.~\ref{LambaSigma}(b). The $\Sigma^{0}$ signal region is defined as $|M_{\gamma\Lambda}-m_{\Sigma^0}|<0.010$~GeV/$c^{2}$, while the sideband regions are defined as [1.151, 1.172]~GeV/$c^{2}$ and [1.213, 1.234]~GeV/$c^{2}$ as indicated by the dashed arrows in Fig.~\ref{LambaSigma}(b).

\begin{figure}[htbp]
  \centering
  \subfigure{\includegraphics[width=7.5cm,height=5.0cm]{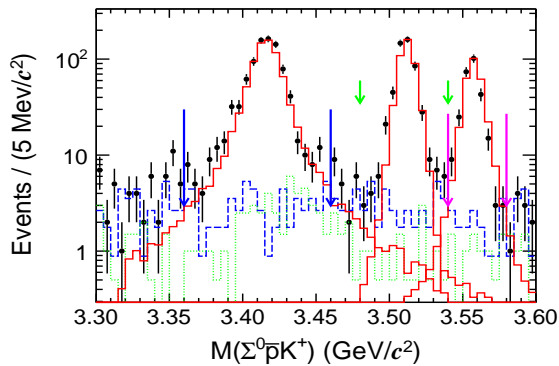}}
  {\caption{The distribution of the $\Sigma^{0}\bar{p}K^{+}$ invariant mass in the region of the $\chi_{cJ}$ states. The dots with error bars are data, the solid histogram is the $\chi_{cJ}$ line shape from MC simulations, the histogram with the dashed line is the background contribution from the inclusive MC sample, where the signal MC simulations and inclusive MC sample have been normalized to the data luminosity. The histogram with the dot line is the normalized $\Sigma^{0}$ sideband, and the solid arrows indicate the $\chi_{c0}$, $\chi_{c1}$, and $\chi_{c2}$ signal regions.}
  \label{chicJm}}
\end{figure}

 The $\Sigma^{0}\bar{p}K^{+}$ invariant mass distribution after application of all selection conditions is shown in Fig.~\ref{chicJm}, where clear $\chi_{c0}$, $\chi_{c1}$, and $\chi_{c2}$ signals are observed. The signal MC simulation also shown in Fig.~\ref{chicJm} agrees with the data very well.

 The $\psi(3686)$ inclusive MC sample is used to study possible peaking backgrounds. Applying the same requirements as the data, the two main remaining background channels involve either $\psi(3686)\to K^{*+}\bar{p}\Lambda$ with $K^{*+}\to K^{+}\pi^{0}$~$(\pi^{0}\to\gamma\gamma$) decays or belong to the peaking background channel $\psi(3686)\to\gamma\chi_{cJ} \to \gamma K^{+}\bar{p}\Lambda$~$(\Lambda\rightarrow p\pi^{-})$ that is missing the intermediate $\Sigma^0$ decay. Other small backgrounds are smoothly distributed below the $\chi_{cJ}$ signal region. These backgrounds can be estimated by the $\Sigma^{0}$ sideband events normalized to the background level below the $\Sigma^{0}$ signal peak. The normalized sideband events are shown as the histogram with the dot line in Fig.~\ref{chicJm}.

\section{\boldmath Measurement of $\mathcal{B}(\chi_{cJ}\to \Sigma^{0}\bar{p}K^{+}+{\rm c.c.})$}

\begin{figure}[htbp]
 \centering
 \subfigure{\includegraphics[width=7.5cm,height=5.5cm]{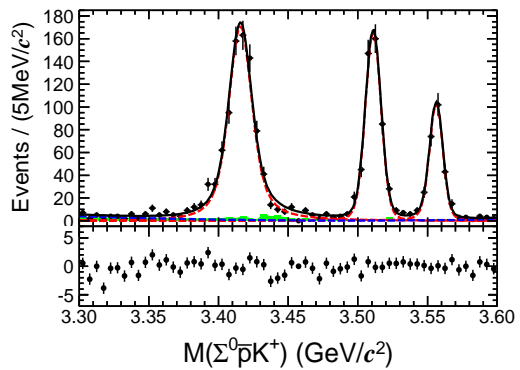}}
 {\caption{Fit to the $M(\Sigma^0\bar{p}K^{+})$ spectrum. Dots with error bars correspond to the data, the black solid curve shows the fit result, the red dashed lines are the signal shapes of the $\chi_{cJ}$ states, the green shaded histogram is the normalized $\Sigma^{0}$ sideband contribution, and the blue dashed line is the continuum background.}
 \label{fitchicJ}}
\end{figure}

 The result of an unbinned maximum likelihood fit to the $M_{\Sigma^{0}\bar{p}K^{+}}$ distribution is shown in Fig.~\ref{fitchicJ}. Here, we fit $\sum_{J}(N_{1,J}\cdot f^{J}_{\rm signal}) + \sum_{J}(N_{2,J}\cdot f^{J}_{\rm peakbkg})+ N_3 \cdot f_{\rm flatbkg}$, where $f^{J}_{\rm signal}$ is the probability density function describing the $\chi_{cJ}$ resonances, $f^{J}_{\rm peakbkg}$ is the normalized shape of the $\Sigma^{0}$ sidebands, and $f_{\rm flatbkg}$ is given by a second-order polynomial. The line shape of each resonance $f_{\rm signal}$ is modeled with the same formula $ BW(M)\cdot E_{\gamma}^{3}\cdot D(E_{\gamma})$ as in Ref.~\cite{isos}, where $M$ is the $\Sigma^{0}\bar{p}K^{+}$ invariant mass, $BW(M)=\frac{1}{(M-m_{\chi_{cJ}})^{2}+\left(\frac{\Gamma_{\chi_{cJ}}}{2}\right)^2}$ is the Breit-Wigner function, $\Gamma_{\chi_{cJ}}$ is the width of the corresponding $\chi _{cJ}$, $E_{\gamma}=\frac{m_{\psi(3686)}^{2}-M^2}{2m_{\psi(3686)}}$ is the energy of the transition photon in the rest frame of the $\psi(3686)$, and $D(E_{\gamma})$ is the damping factor which suppresses the divergent tail due to the $E_{\gamma}^{3}$ dependence of $f^{J}_{\rm signal}$. It is described by $\exp(-E_{\gamma}^{2}/8\beta^{2})$, where $\beta=(65.0\pm2.5)$~MeV was measured by the CLEO experiment~\cite{CLEO}. The signal shapes are convolved with Gaussian functions to account for the mass resolution.

 The parameters $N_{1,J}$, $N_3$ and two coefficients of the polynomial are taken as the free parameters in the fit, while $N_{2,J}$ is fixed to the number of the normalized $\Sigma^{0}$ sideband events. In the description of $f^{J}_{\rm signal}$, the masses and widths of the $\chi_{cJ}$ states are fixed to the PDG values. The Gaussian resolution parameters in the region of the three $\chi_{cJ}$ states are also free parameters, and are found to be $5.7$, $5.1$, and $4.1$~MeV/$c^{2}$ for $\chi_{c0}$, $\chi_{c1}$, and $\chi_{c2}$, respectively. The yields of signal events of all three $\chi_{cJ}\to\Sigma^{0}\bar{p}K^{+}$ decays are listed in Table~\ref{result}.

\begin{figure*}[htbp]
\centering
\subfigure{\includegraphics[width=5.0cm,height=4.0cm]{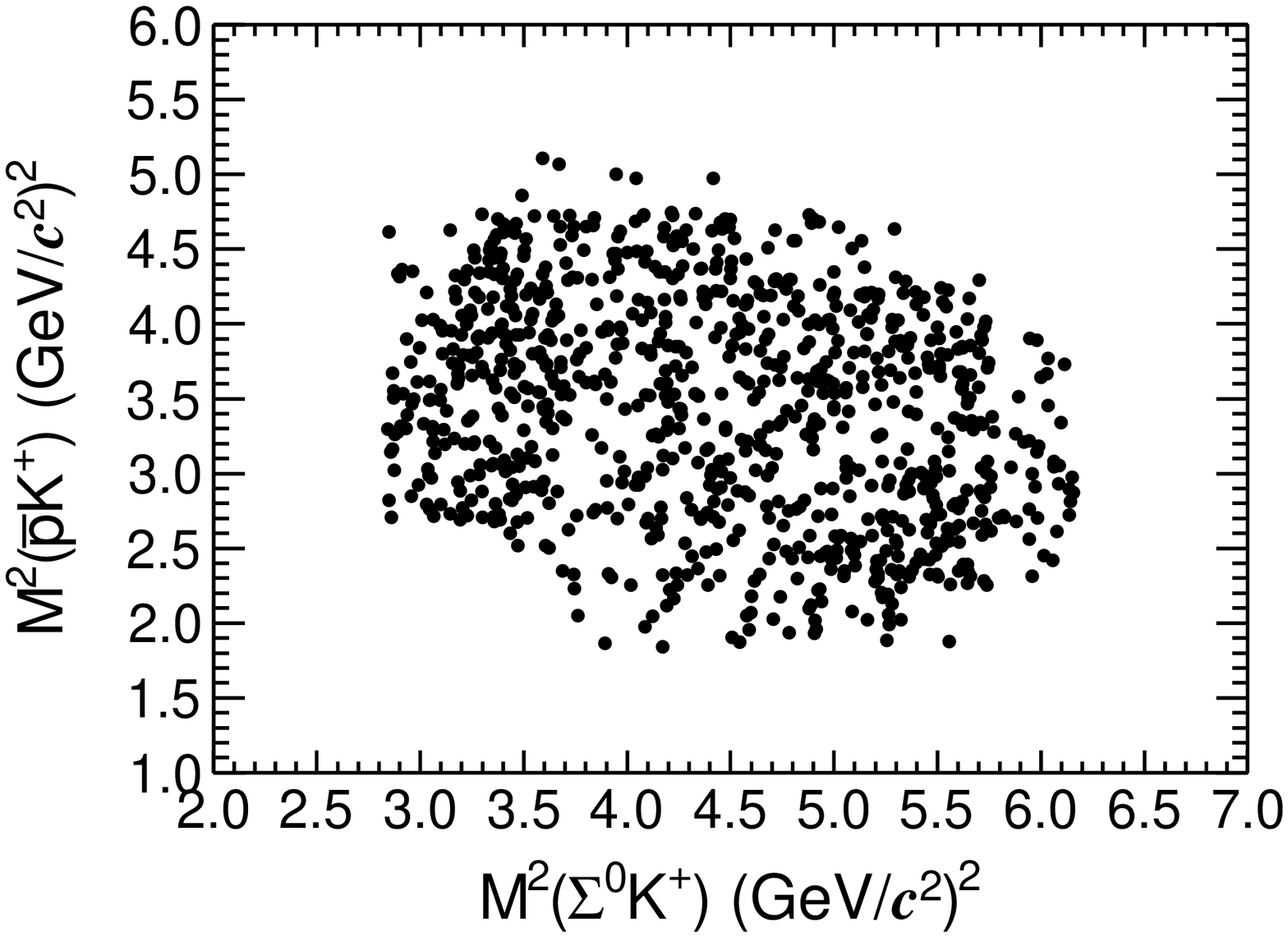}\put(-25,98){(a)}}
\subfigure{\includegraphics[width=5.0cm,height=4.0cm]{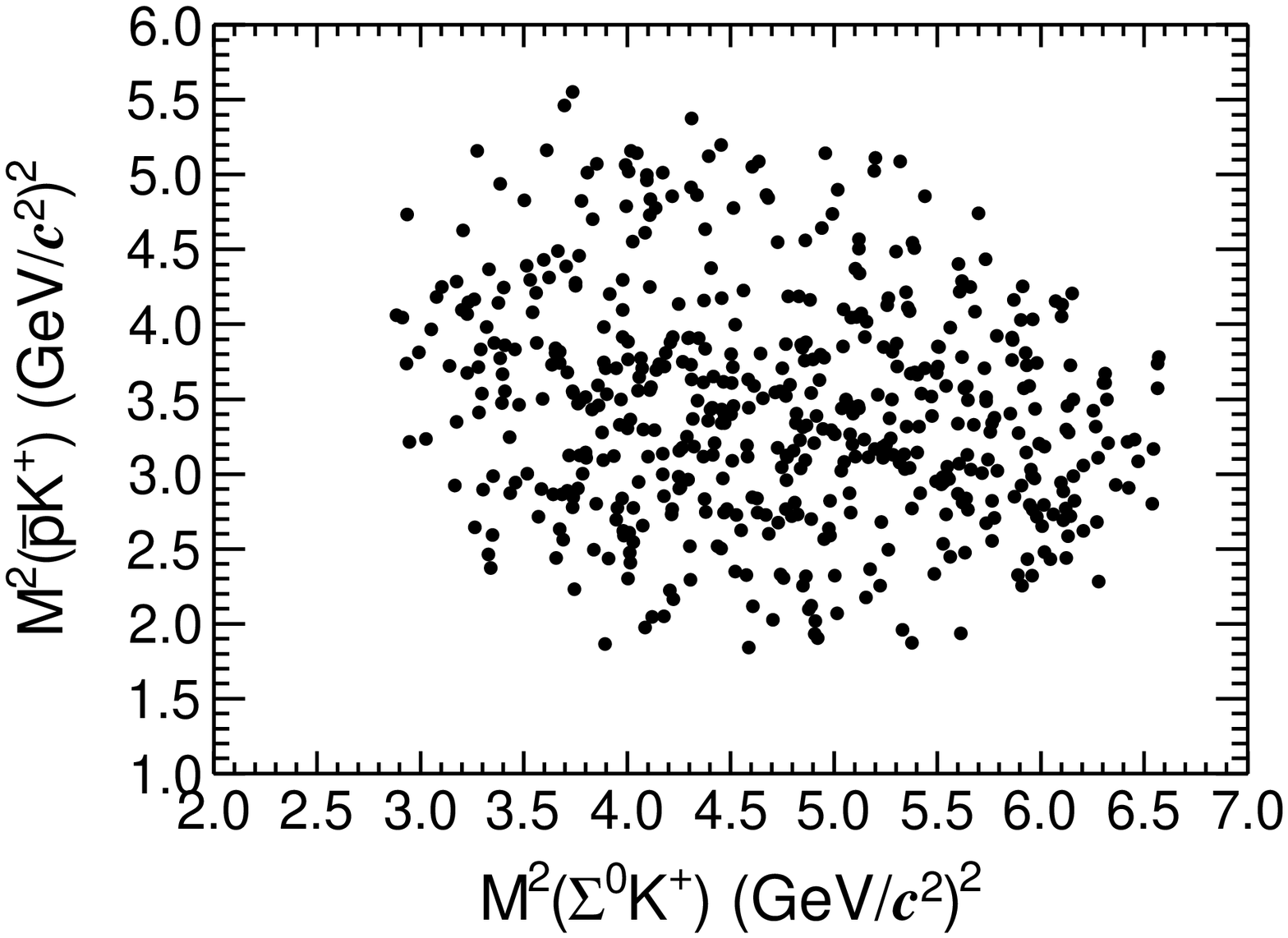}\put(-25,98){(b)}}
\subfigure{\includegraphics[width=5.0cm,height=4.0cm]{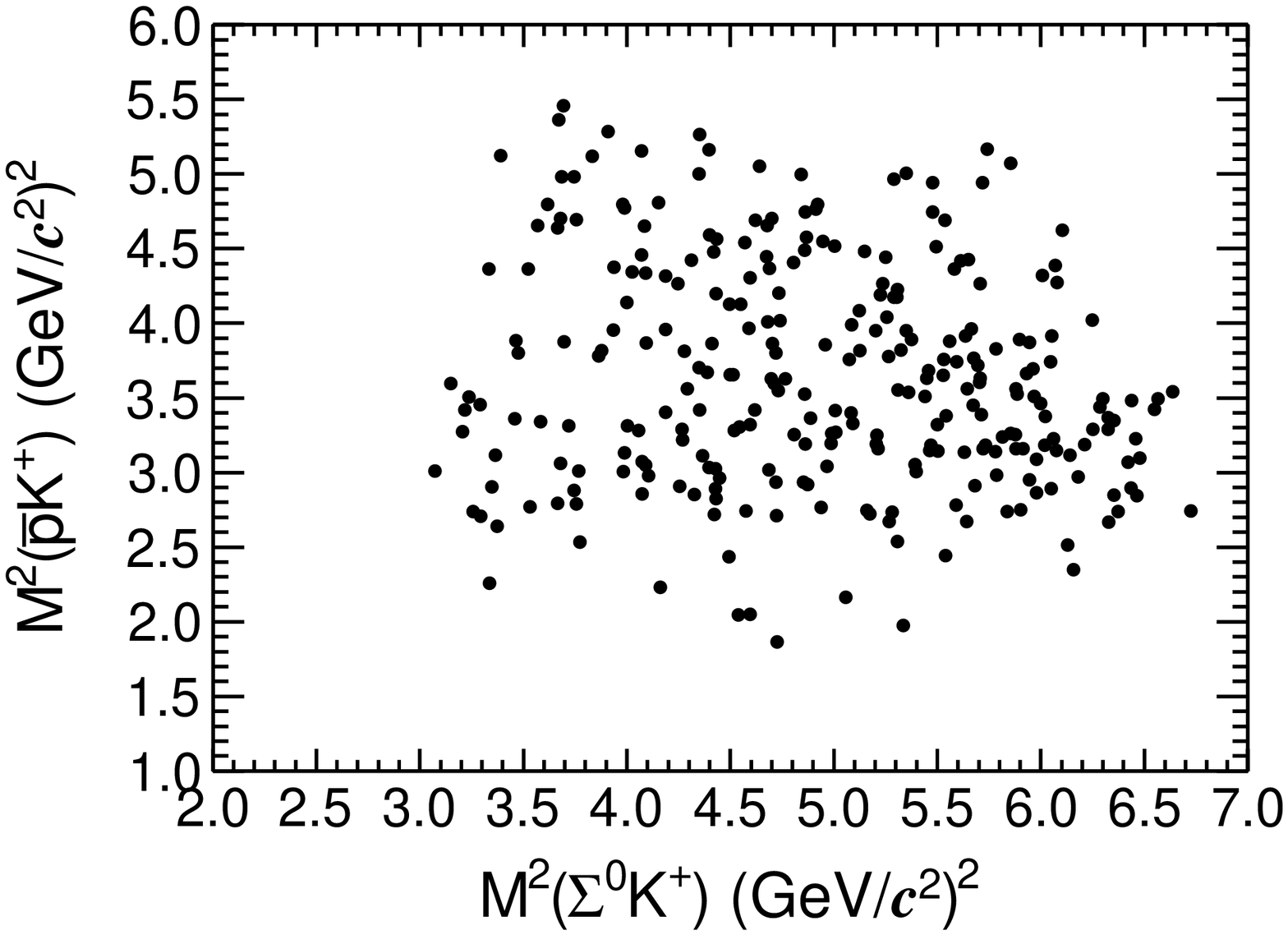}\put(-25,98){(c)}}

\subfigure{\includegraphics[width=5.0cm,height=4.0cm]{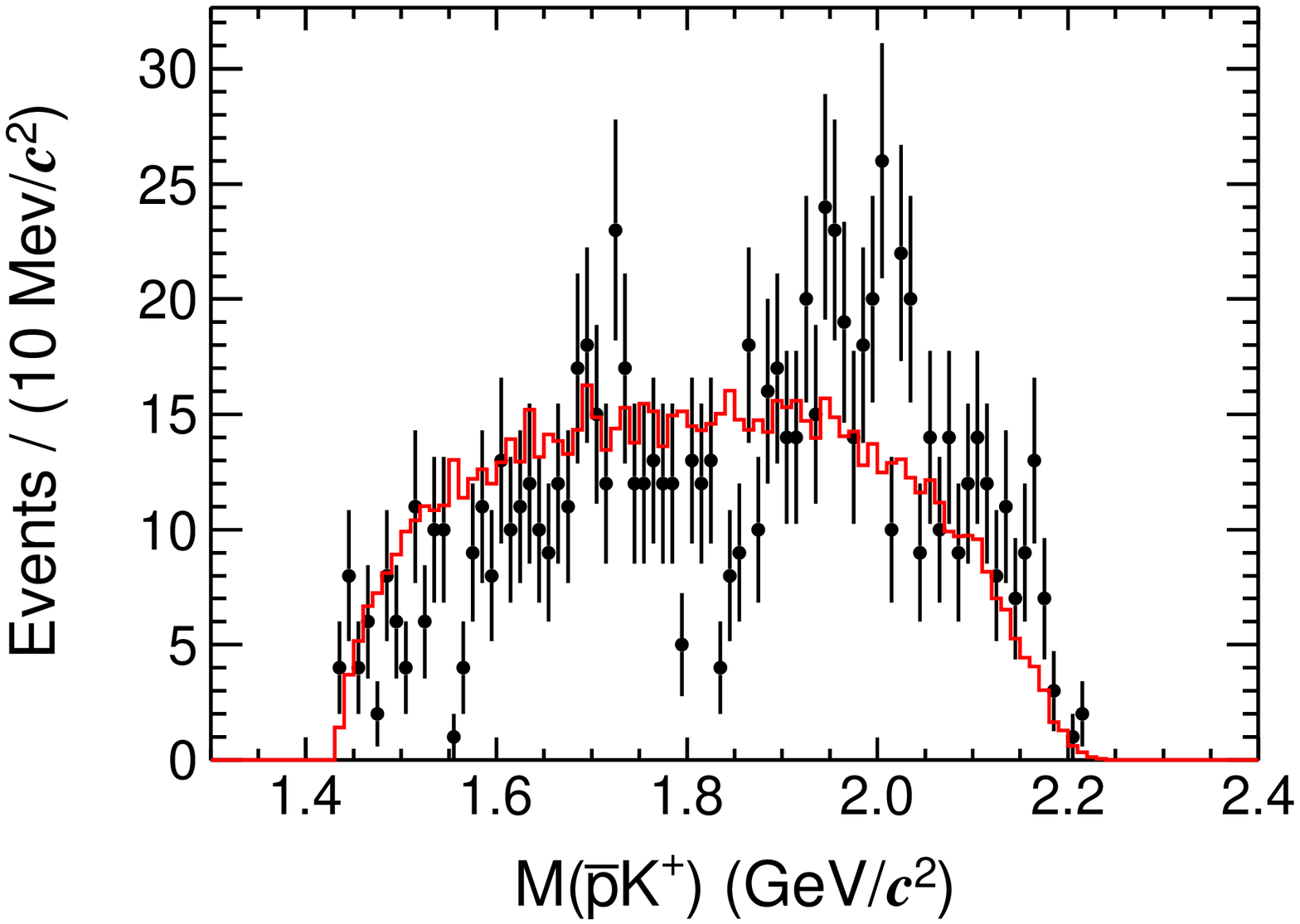}\put(-25,98){(d)}}
\subfigure{\includegraphics[width=5.0cm,height=4.0cm]{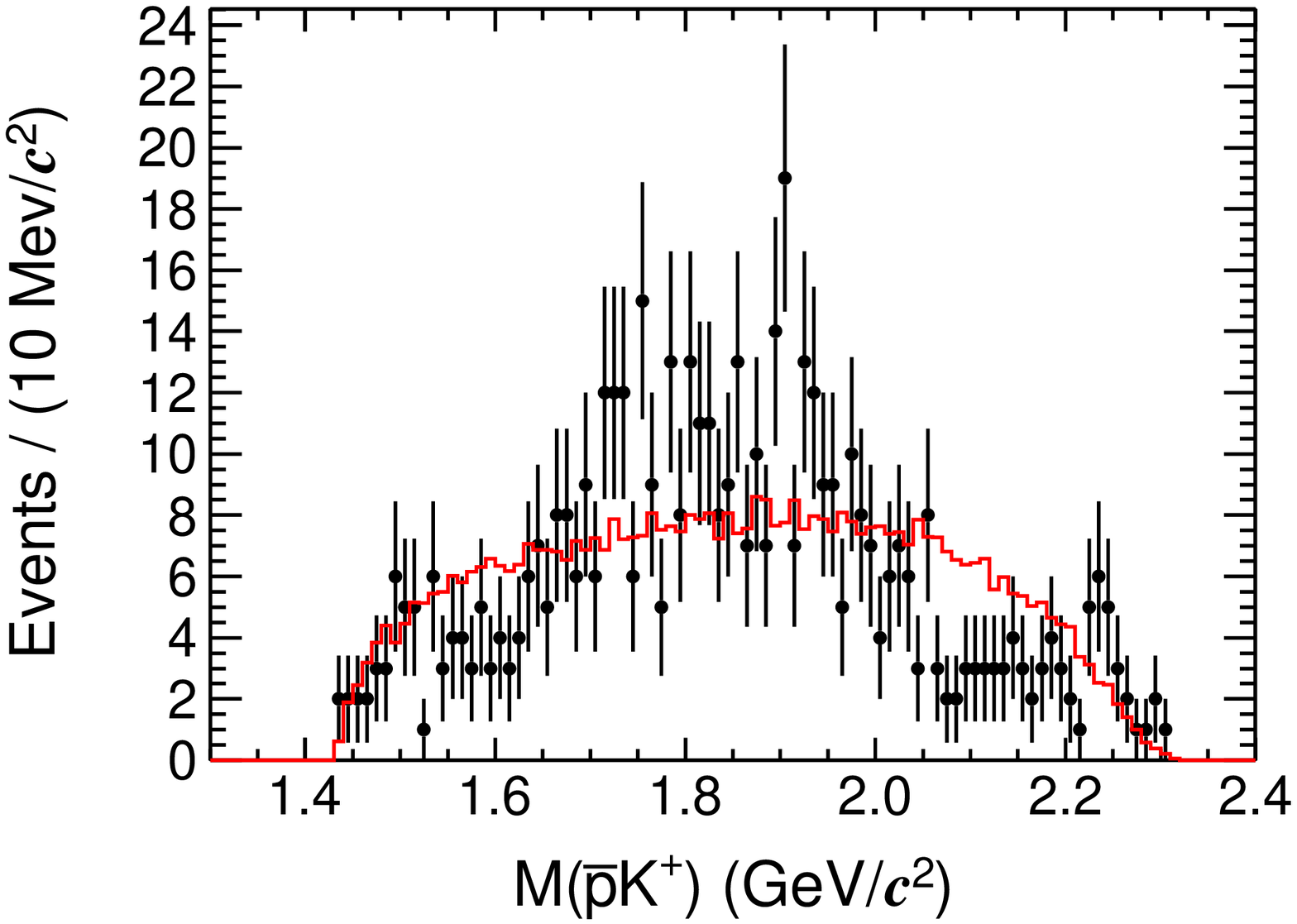}\put(-25,98){(e)}}
\subfigure{\includegraphics[width=5.0cm,height=4.0cm]{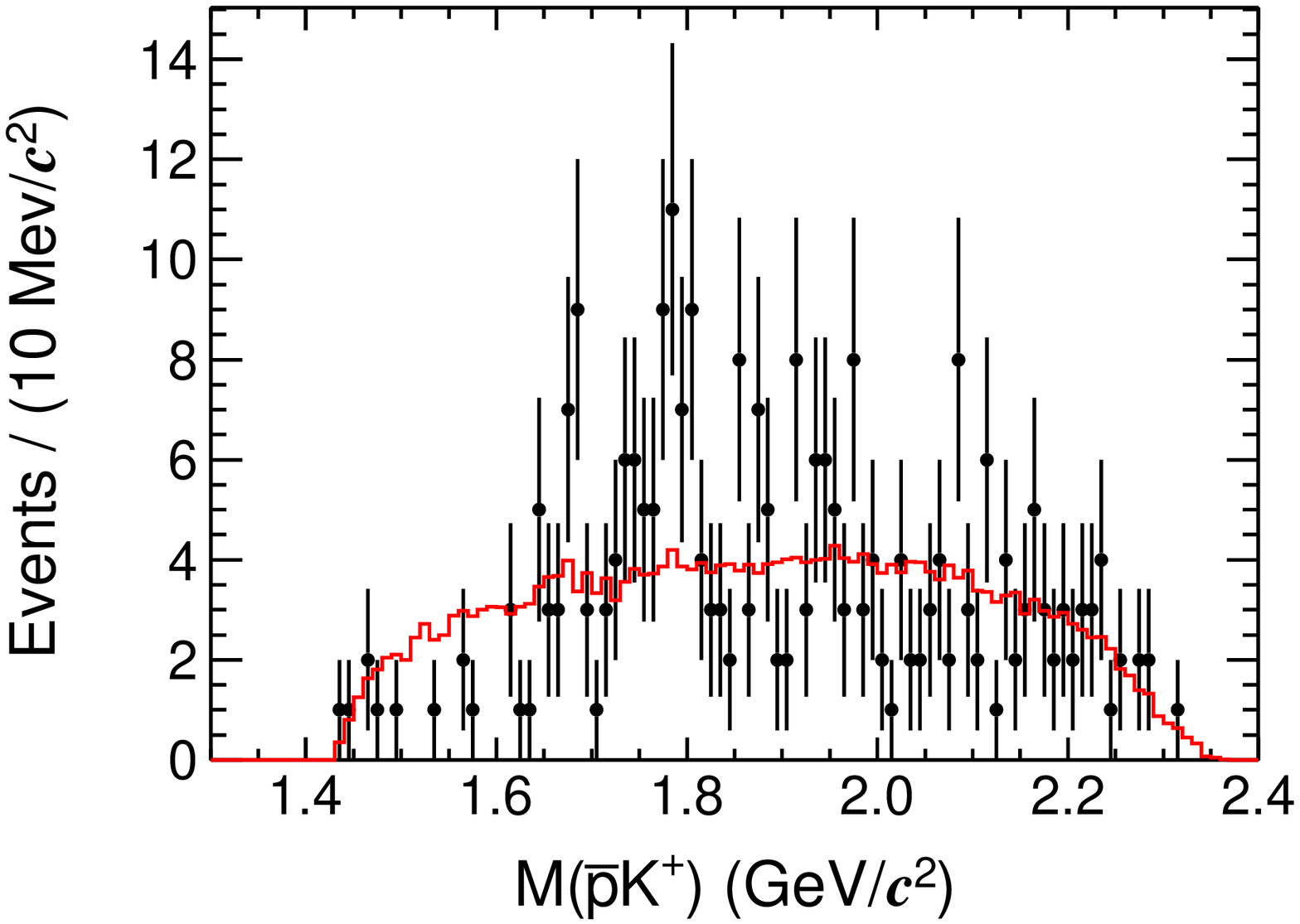}\put(-25,98){(f)}}

\subfigure{\includegraphics[width=5.0cm,height=4.0cm]{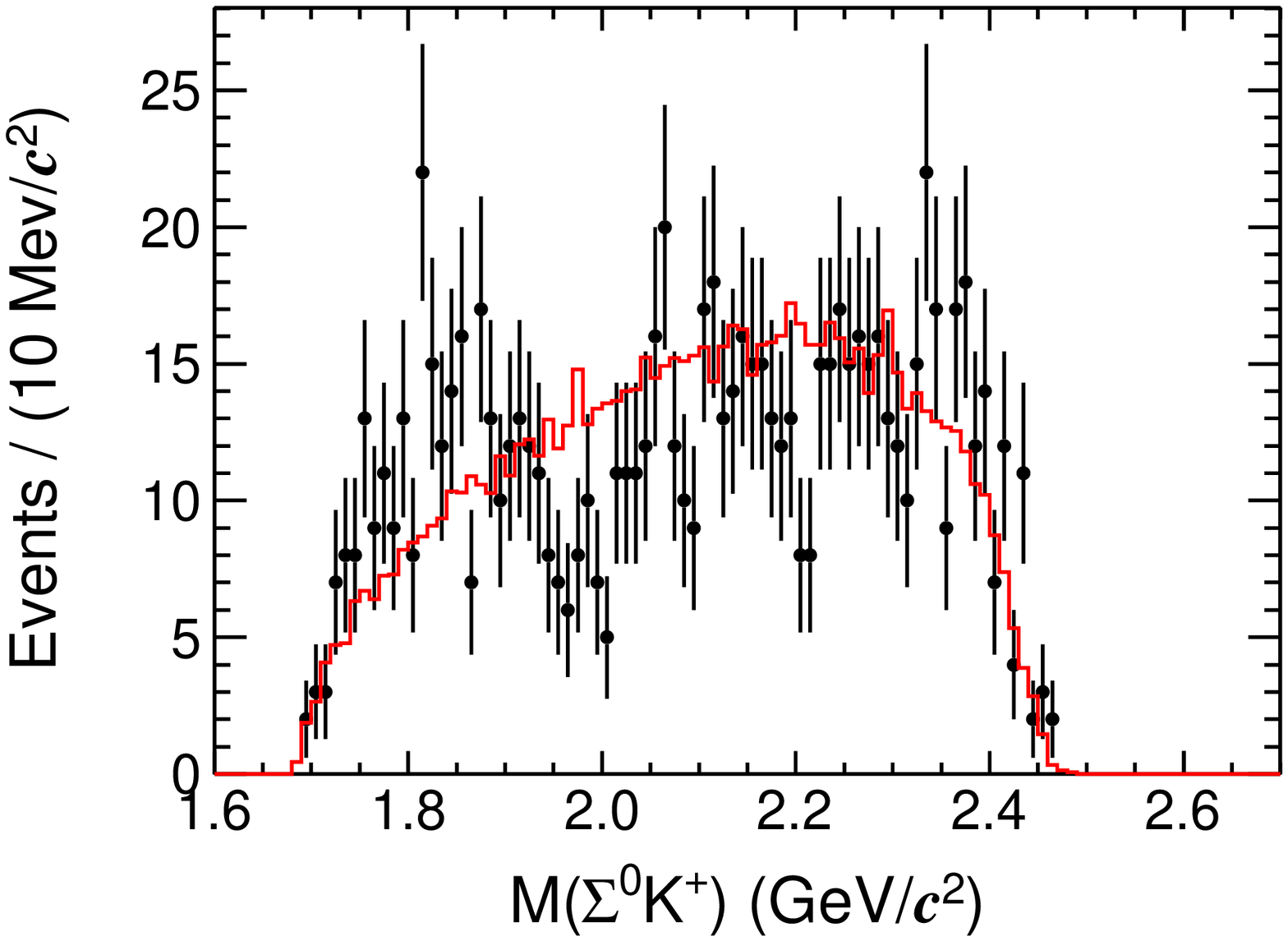}\put(-25,98){(g)}}
\subfigure{\includegraphics[width=5.0cm,height=4.0cm]{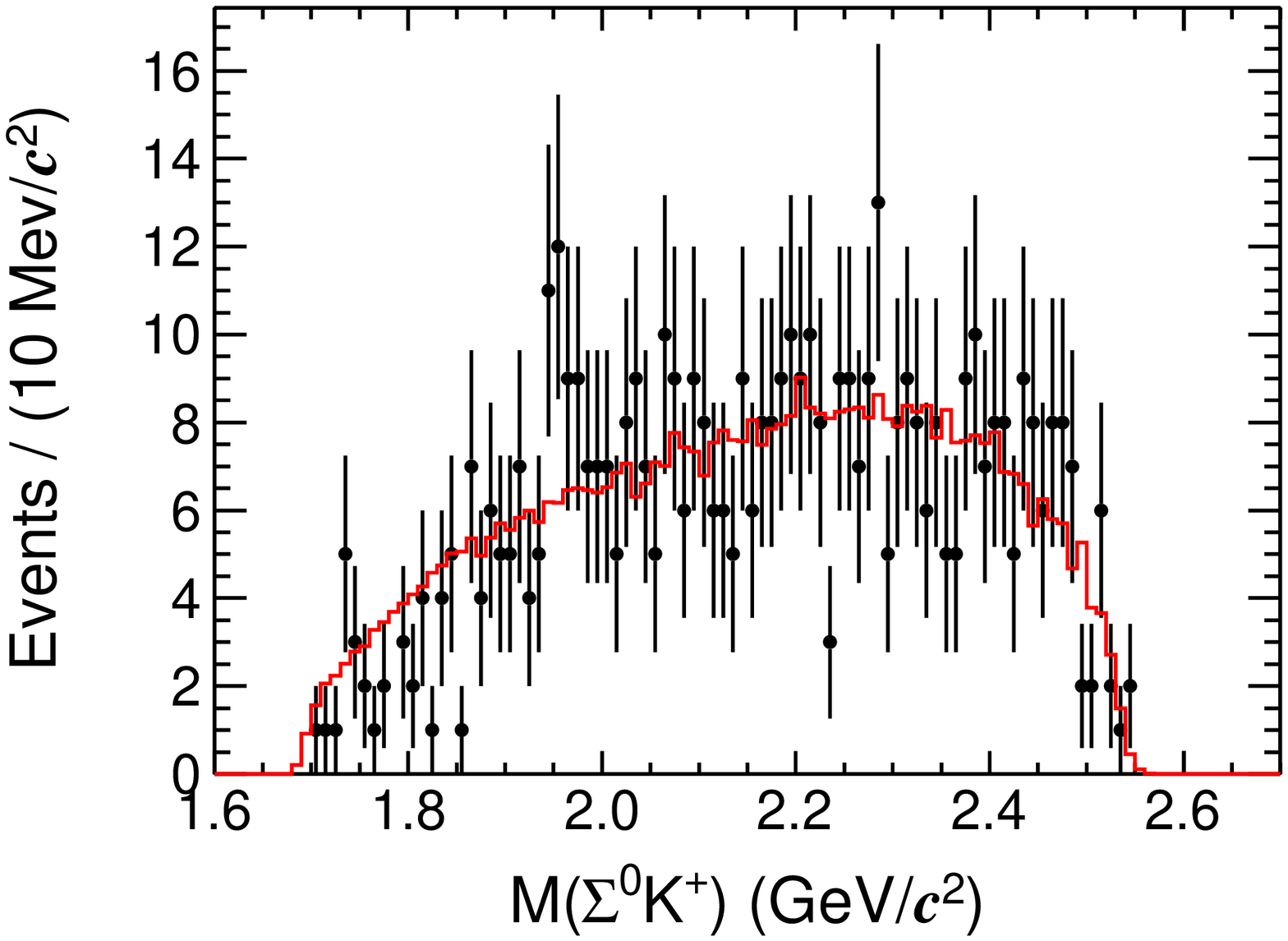}\put(-25,98){(h)}}
\subfigure{\includegraphics[width=5.0cm,height=4.0cm]{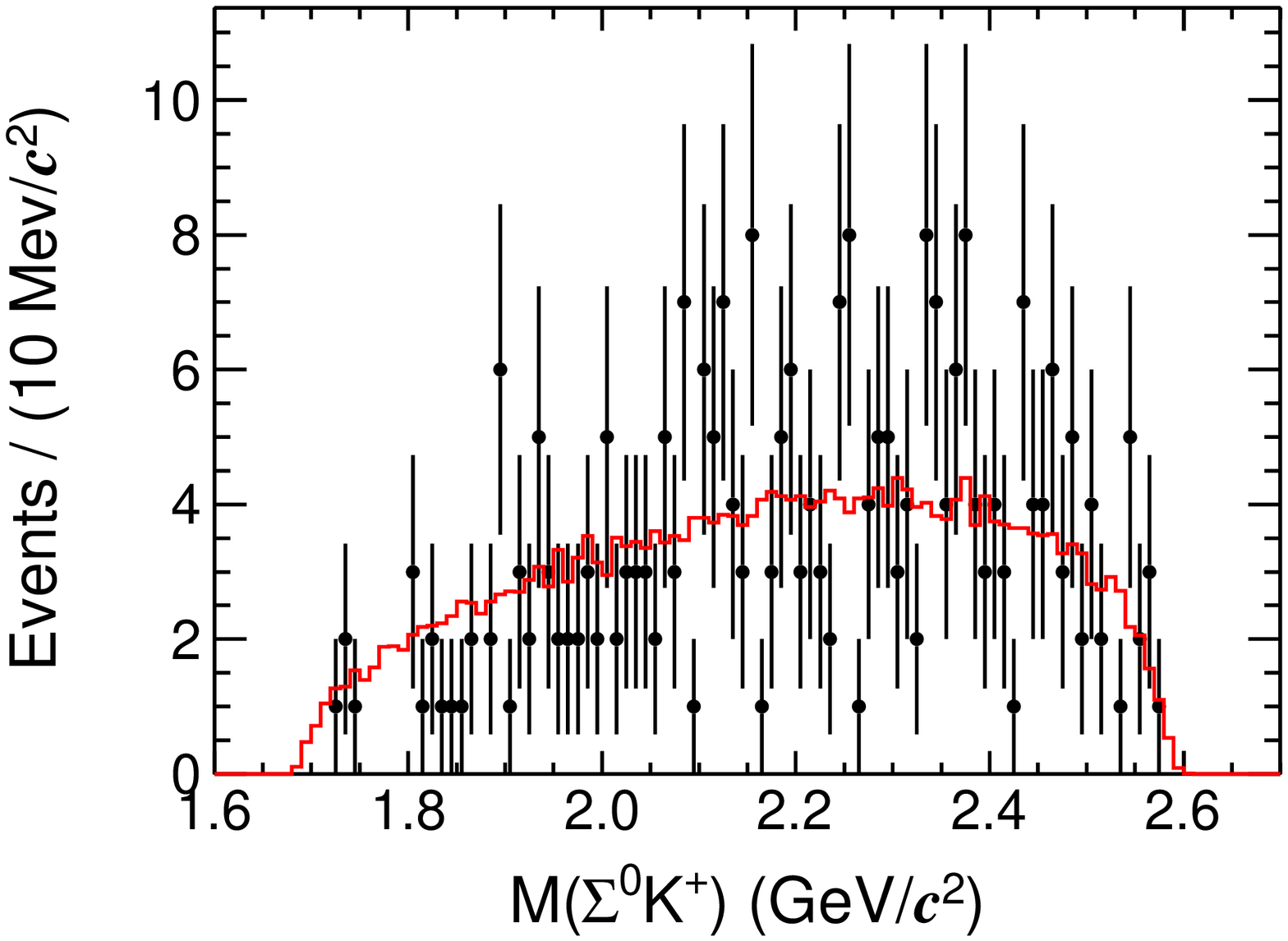}\put(-25,98){(i)}}

\subfigure{\includegraphics[width=5.0cm,height=4.0cm]{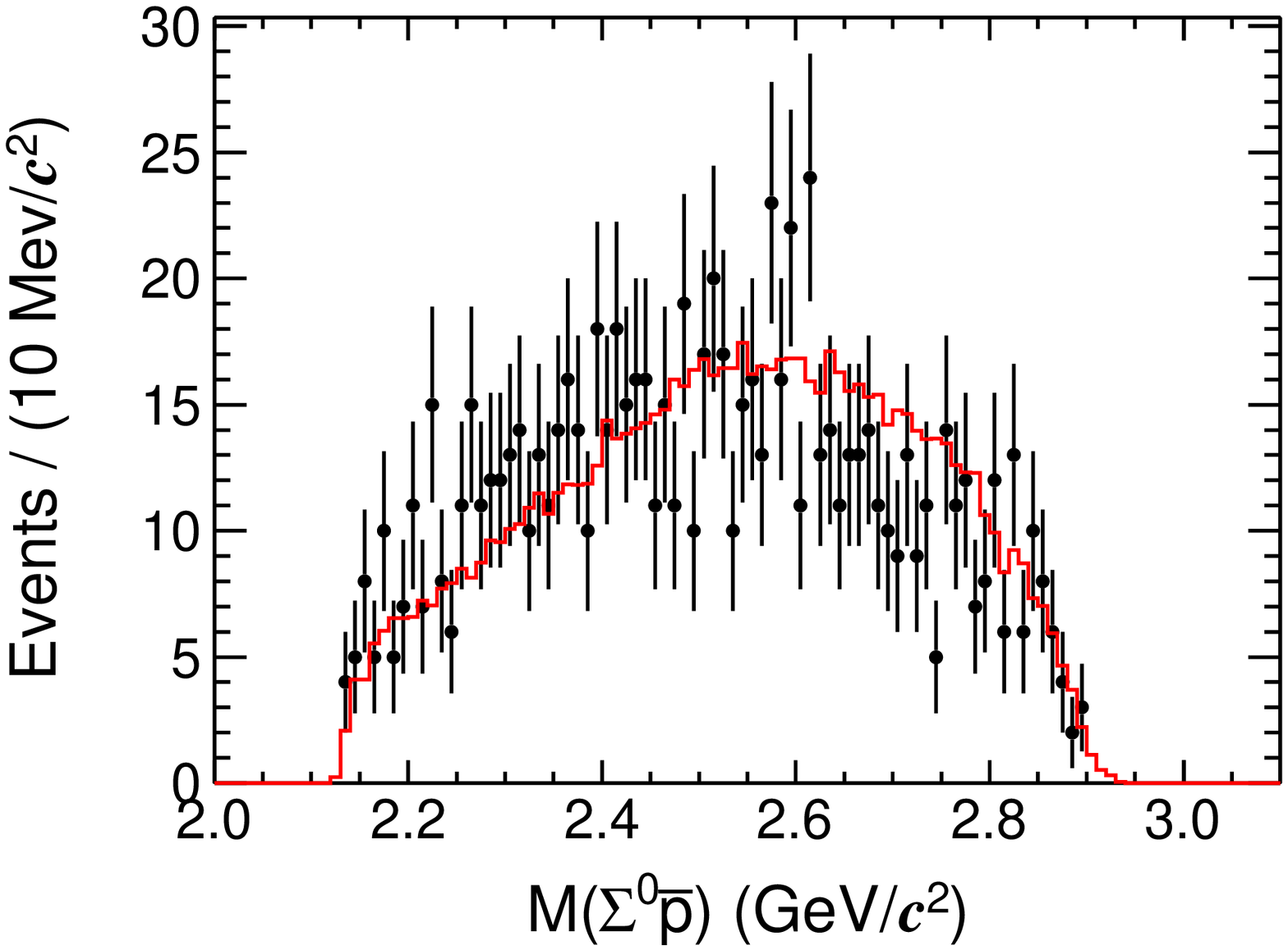}\put(-25,98){(j)}}
\subfigure{\includegraphics[width=5.0cm,height=4.0cm]{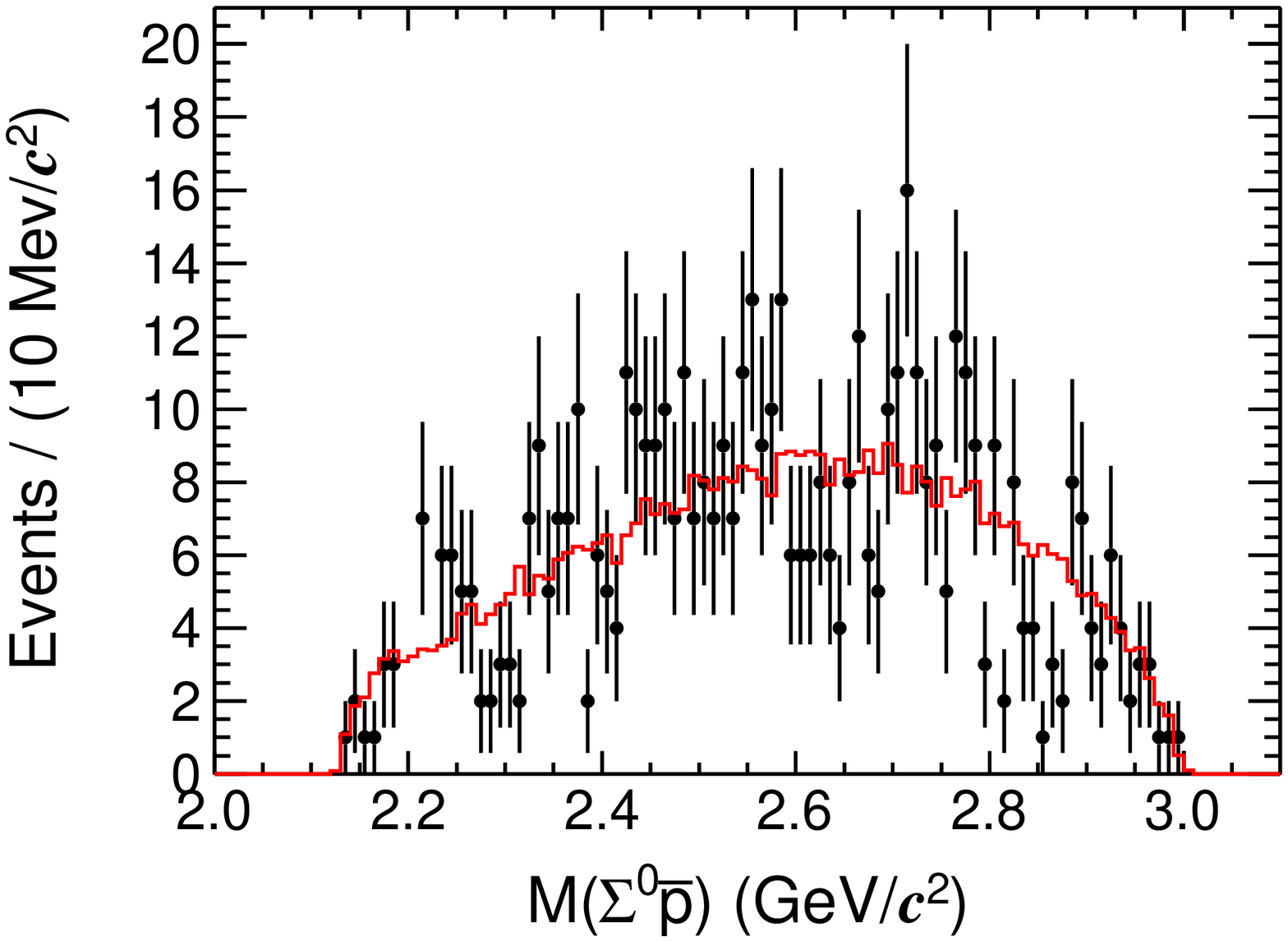}\put(-25,98){(k)}}
\subfigure{\includegraphics[width=5.0cm,height=4.0cm]{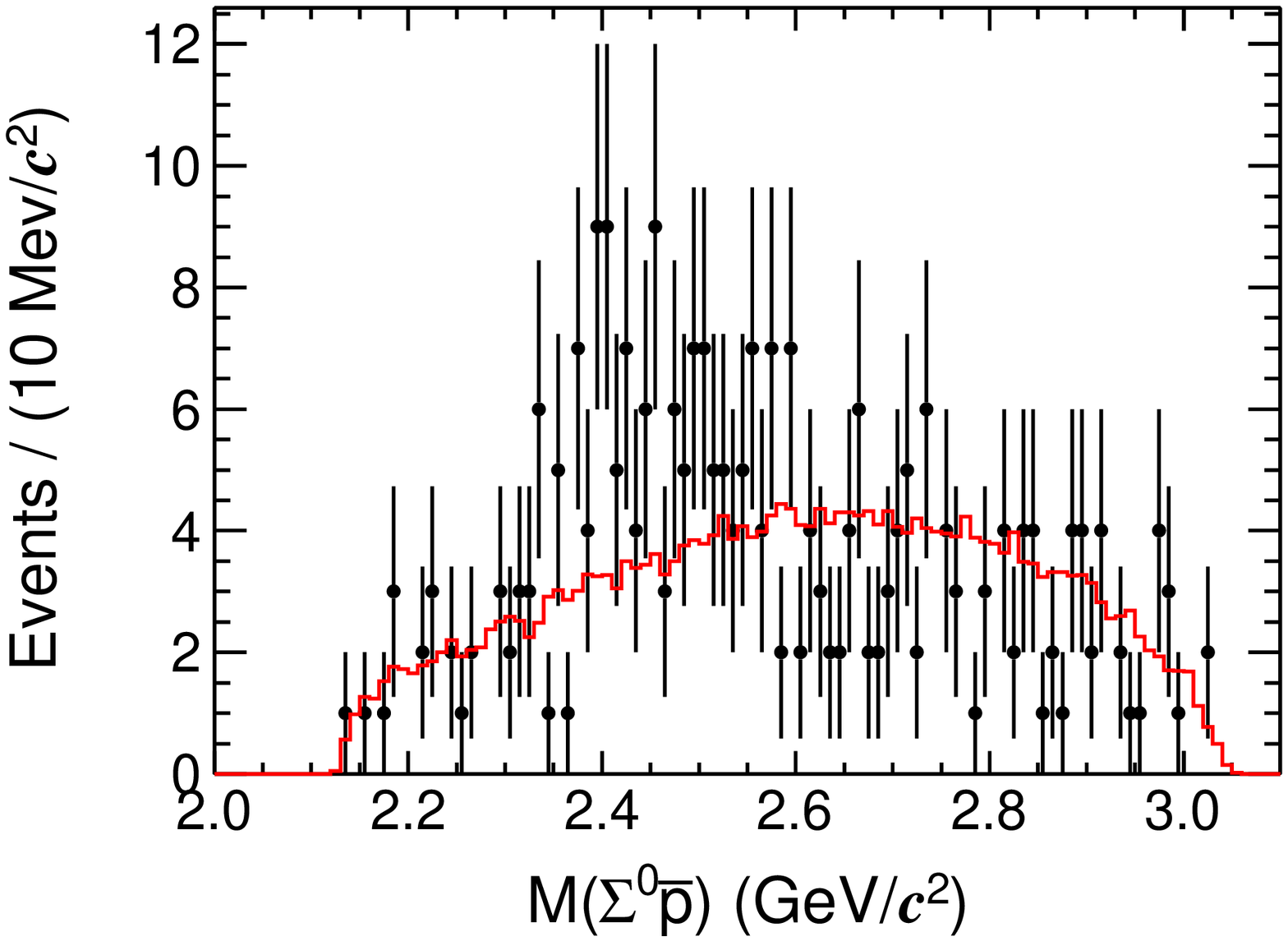}\put(-25,98){(l)}}
{\caption{Dalitz plots and one-dimensional projections of $\chi_{cJ}\to \Sigma^{0}\bar{p}K^{+}+{\rm c.c.}~(J=0, 1, 2)$. The left column (a, d, g, j) is for $\chi_{c0}$, the middle column (b, e, h, k) is for $\chi_{c1}$, and the right column (c, f, i, l) is for $\chi_{c2}$. Dots with error bars are the data, the histograms with solid lines represent phase-space MC simulations. }
\label{Dalitzplot}}
\end{figure*}

 Dalitz plots and the one dimensional projections of $\chi_{cJ}\to\Sigma^{0}\bar{p}K^{+}$ events are shown in the left, middle and right columns of Fig.~\ref{Dalitzplot} for the $\chi_{c0}$, $\chi_{c1}$, and $\chi_{c2}$ signal regions, respectively, together with the distributions of MC simulated signal events based on a pure phase-space decay model.

 For $\bar{p}K^{+}$ mass spectra of the data, it seems there are two structures around 1.7 and 2.0~GeV/$c^{2}$ for $\chi_{c0}$ decays, they are likely $\bar{\Sigma}(1750)^{0}$ and $\bar{\Sigma}(1940)^{0}$. There seems to be two structures around 1.9~GeV/$c^2$ for $\chi_{c1}$ decays and around 1.8~GeV/$c^2$ for $\chi_{c2}$ decays. For $\Sigma^{0}K^{+}$ mass spectra, it seems there is a jump around 1.8~GeV/$c^2$ and a dip around 2.0~GeV/$c^2$ for $\chi_{c0}$ decays, the jump may be $N(1880)$ with $J^P=\frac{1}{2}^{+}$ or $N(1895)$ with $J^P=\frac{1}{2}^{-}$. There is an indication around 1.95~GeV/$c^2$ for $\chi_{c1}$ decays, which may be $N(1900)$ with $J^P=\frac{3}{2}^{+}$. There is no evident structure for $\chi_{c2}$ decays. For $\Sigma^{0}\bar{p}$ mass spectra, the data are consistent with the phase-space MC shapes, there is no evident structure for $\chi_{c0}$, $\chi_{c1}$ and $\chi_{c2}$ decays. The mass distributions of two-body subsystems of the data are not completely consistent with the phase-space MC simulations, but it is difficult to draw any conclusions to them due to present limited statistics.

 The differences between data and MC simulation indicate that these signal MC events cannot be used to calculate the selection efficiency directly. Instead, the detection efficiency is obtained by weighting the simulated Dalitz plot distribution with the distribution from data. We divide the Dalitz plots of $M_{\bar{p}K^{+}}^2$ versus $M_{\Sigma^{0}K^{+}}^2$ into $12\times12$, $8\times7$, and $6\times8$ bins in the $\chi_{c0}$, $\chi_{c1}$, and $\chi_{c2}$ regions, respectively. First, we obtain the weight factor $\omega_i$ in each bin as the ratio between the Dalitz plot distribution of data and the normalized signal MC sample. In a second step, $\omega_i$ is used to correct the Dalitz distributions of both the generated and reconstructed MC simulations. Finally, we determine the corrected detection efficiency as the ratio between the sum of event weights in reconstructed and generated MC. The results are listed in Table~\ref{result}.

The branching fractions for $\chi_{cJ}\to\Sigma^{0}\bar{p}K^{+}+{\rm c.c.}$ are calculated using
\begin{equation}
  \begin{aligned}
      \mathcal{B}(\chi_{cJ}\to \Sigma^{0}\bar{p}K^{+}+{\rm c.c.})
      =\frac{N^{\rm obs}}{N_{\psi(3686)}\cdot\epsilon\cdot \prod_{j}\mathcal{B}_{j}},
  \end{aligned}
\end{equation}
where $N^{\rm obs}$ is the number of signal events obtained from the fit, $N_{\psi(3686)}$ is the total number of $\psi(3686)$ events, $\epsilon$ is the corresponding detection efficiency as listed in Table~\ref{result}, and
$\prod_{j}\mathcal{B}_{j} = \mathcal{B}(\psi(3686)\to
\gamma\chi_{cJ})\times\mathcal{B}(\Sigma^{0}\to \gamma\Lambda)\times\mathcal{B}(\Lambda\to p\pi^{-})$ is the product branching fraction with individual values taken from the PDG~\cite{PDG}. The results for $\chi_{cJ}\to \Sigma^{0}\bar{p}K^{+}+{\rm c.c.}~(J=0, 1, 2)$ are listed in Table~\ref{result}.

\begin{table*}[htp]
     \centering
    {\caption{Summary of the number of fitted  signal events ($N^{\rm obs}$), detection efficiency ($\epsilon$), and branching fraction $\mathcal{B}(\chi_{cJ}\to\Sigma^{0}\bar{p}K^{+}+{\rm c.c.})$, where the first uncertainty is statistical and the second one is systematic.}
    \label{result}}
    \begin{tabular}{l c c c}
    \hline \hline
    Mode & $N^{\rm obs}$ & $\epsilon ~(\%)$ & $\mathcal{B}(\chi_{cJ}\to\Sigma^{0}\bar{p}K^{+}+{\rm c.c.})(10^{-4})$ \\
    \hline
    $\chi_{c0}\to \Sigma^{0}\bar{p}K^{+}$ & ~~ $871 \pm 34$ ~~& ~~$10.25\pm 0.05$~~& ~~$3.03\pm 0.12\pm 0.15$\\
    $\chi_{c1}\to \Sigma^{0}\bar{p}K^{+}$ & ~~ $493 \pm 24$ ~~& ~~$12.12\pm 0.05$~~& ~~$1.46\pm 0.07\pm 0.07$\\
    $\chi_{c2}\to \Sigma^{0}\bar{p}K^{+}$ & ~~ $271 \pm 18$ ~~& ~~$10.90\pm 0.05$~~& ~~$0.91\pm 0.06\pm 0.05$\\
    \hline \hline
    \end{tabular}
\end{table*}

\section{Systematic Uncertainties}

The systematic uncertainties on the measurement of the branching fractions of $\chi_{cJ}\to \Sigma^{0}\bar{p}K^{+}+{\rm c.c.}$ are discussed below.

Using the control samples of $J/\psi\to p\bar{p}\pi^{+}\pi^{-}$ and $J/\psi\to K^{*}\bar{K}$, the difference of tracking efficiencies between MC simulation and data is within 1\% for $\bar{p}$ and $K^{+}$. Therefore, 2\% is taken as the tracking systematic uncertainty.

The $\bar{p}/K^{+}$ PID efficiency is studied using $J/\psi\to p\bar{p}\pi^{+}\pi^{-}$ and $J/\psi\to K^{0}_{S}K^{\pm}\pi^{\pm}$ control samples~\cite{PIDproton,PIDKaon}, with the result being that the PID efficiency for data agrees with that of the MC simulation within 1\% per $\bar{p}/K^{+}$. So 2\% is taken as the systematic uncertainty associated with the PID efficiency.

The photon detection efficiency is studied from a $J/\psi\to\pi^+\pi^-\pi^0$ control sample~\cite{photon}. The efficiency difference between data and MC simulation is about 1\% per photon, so that 2\% is assigned as the systematic uncertainty from the two photons.

In order to determine the uncertainty associated with the secondary vertex fit and the decay length requirement, we determine the efficiency of these selection criteria by comparing the $\Lambda\to p\pi^-$ signal yields with and without those selections for both data and signal MC. From a fit to the $p\pi^{-}$ invariant mass distributions, we find a data-MC difference of 0.7\% that is assigned as the systematic uncertainty. For each track stemming from $\Lambda\to p\pi^-$ decays, the systematic uncertainty from the tracking efficiency is 1.0\% according to an analysis of $J/\psi\to \bar{p} K^{+} \Lambda$~\cite{tracking}. The total uncertainty of the $\Lambda$ reconstruction is 2.1\%.

The uncertainty associated with the 4C kinematic fit comes from a potential inconsistency between data and MC simulation; this difference is reduced by correcting the track helix parameters in the MC simulation, as described in detail in Ref.~\cite{refsmear}. The difference of the efficiency with and without the helix correction is considered as the systematic uncertainty from the kinematic fit.

The uncertainty related to the $\Lambda$ and $\Sigma^{0}$ mass windows is studied by determining the yield of $\Lambda$ ($\Sigma^{0}$) inside the mass windows for both data and signal MC simulation. The difference between data and MC simulation is found to be negligible for $\Lambda$, and to be 0.2\% for $\Sigma^{0}$.

In the weighting procedure, the Dalitz plots were divided into $12\times12$, $8\times7$ and $6\times8$ bins in order to calculate the event-weights used in the efficiency determination. We repeat this procedure with different bin configurations. The maximum difference between the nominal binning and the alternate configuration is taken as the weighting related uncertainty listed in Table~\ref{summary_of_syserr}.
The statistical uncertainty of the efficiency is determined directly from MC simulations and amounts to less than 0.5\%.

The systematic uncertainty related to the fitting procedure includes multiple sources.
Concerning the signal line shape, the damping factor is changed from $\exp(-E_{\gamma}^{2}/8\beta^{2})$ as used by CLEO to $\frac{E_{0}^{2}}{E_{0}E_{\gamma}+(E_{0}-E_{\gamma})^2}$ as used by KEDR~\cite{KEDR}. The resulting differences in the fit are assigned as the systematic uncertainties. In addition, the fit range is varied from [3.30, 3.60]~GeV/$c^{2}$ to [3.30, 3.65]~GeV/$c^{2}$ and [3.25, 3.60]~GeV/$c^{2}$ and the maximum differences in the fitted yields are considered as the associated systematic uncertainties. Regarding the peaking background contributions, the $\Sigma^{0}$ sideband ranges were changed from [1.151, 1.172], [1.213,1.234]~GeV/$c^{2}$ to [1.153, 1.174], [1.211, 1.232]~GeV/$c^{2}$ and the difference in signal yields is taken as the systematic uncertainty. With regard to non-$\chi_{cJ}$ backgrounds, the fit function is changed from a second to a third order polynomial in the fit to the $\Sigma^0\bar{p}K^{+}$ invariant mass distribution and the difference between the two fits is taken as the systematic uncertainty.

The systematic uncertainties due to the branching fractions of
$\psi(3686) \to \gamma\chi_{c0}~(\chi_{c1}, ~\chi_{c2})$, and $\Lambda\to p\pi^{-}$, are 2.0\% (2.5\%, 2.1\%), and 0.8\% according to the PDG~\cite{PDG}. For the $\Sigma^{0}\to \gamma\Lambda$ decay, no uncertainty is given in the PDG.

The number of $\psi(3686)$ events is determined to be $(448.1\pm2.9)\times 10^6$~ from inclusive hadronic events~\cite{psidata}, thus the uncertainty is 0.6\%.

All systematic uncertainty contributions discussed above are summarized in Table~\ref{summary_of_syserr}. The total systematic uncertainty for each $\chi_{cJ}$ decay is obtained by adding all contributions in quadrature.

  \begin{table}[htbp]
    \centering
    {\caption {Summary of systematic uncertainty sources and their contributions (in \%). }
    \label{summary_of_syserr}}
    \begin{tabular}{lccc}
      \hline \hline
       Source &  $\mathcal{B}(\chi_{c0} ) $& $\mathcal{B}(\chi_{c1} )$&$\mathcal{B}(\chi_{c2} )$\\
       \hline
       Tracking                 & 2.0 & 2.0& 2.0  \\
       PID                      & 2.0 & 2.0& 2.0  \\
       Photon detection         & 2.0 & 2.0& 2.0  \\
       $\Lambda$ reconstruction & 2.1 & 2.1& 2.1  \\
       4C  kinematic fit        & 0.7 & 0.1& 1.0  \\
      $\Lambda$  mass window & $\cdot$ $\cdot$ $\cdot$  & $\cdot$ $\cdot$ $\cdot$& $\cdot$ $\cdot$ $\cdot$\\
      $\Sigma^{0}$  mass window & 0.2 & 0.2& 0.2  \\
      Weighting procedure       & 1.2 & 0.3& 1.0  \\
      MC statistics             & 0.5 & 0.5& 0.5  \\
      Fitting procedure         & 1.4 & 1.1& 1.0  \\
    Secondary branching fractions&2.2 & 2.6& 2.2  \\
      Number of $\psi(3686)$    & 0.6 &0.6 & 0.6  \\ \hline
      Total                     & 5.1 &5.0 & 5.0  \\
      \hline\hline
    \end{tabular}
  \end{table}

\section{Summary}

 Using the $(448.1\pm2.9)\times$10$^{6}$ $\psi(3686)$ events accumulated with the BESIII detector, the three-body decays of $\chi_{cJ}\to \Sigma^{0}\bar{p}K^{+}+{\rm c.c.}~(J = 0, 1, 2)$ are studied for the first time, and clear $\chi_{cJ}$ signals are observed. The branching fractions of $\chi_{cJ}\to \Sigma^{0}\bar{p}K^{+}+$~c.c. are determined to be $(3.03\pm0.12\stat\pm0.15\syst) \times10^{-4}$, $(1.46\pm0.07\stat\pm0.07\syst)\times10^{-4}$, and $(0.91\pm0.06\stat\pm0.05\syst)\times10^{-4}$ for $J=0$, 1, and 2, respectively.

 Comparing with the isospin conjugate decays of $\chi_{cJ}\to\Sigma^{+}\bar{p}K^{0}_{S}+{\rm c.c.}~(J = 0, 1, 2)$~\cite{isos}, we obtain the ratios of the branching fractions $\frac{\mathcal{B}(\chi_{c0}\to \Sigma^{0}\bar{p}K^{+})}{\mathcal{B}(\chi_{c0}\to \Sigma^{+}\bar{p}K^{0}_{S})} = 0.86 \pm 0.06 \pm 0.06$, $\frac{\mathcal{B}(\chi_{c1}\to \Sigma^{0}\bar{p}K^{+})}{\mathcal{B}(\chi_{c1}\to \Sigma^{+}\bar{p}K^{0}_{S})} = 0.95 \pm 0.08 \pm 0.06$, and $\frac{\mathcal{B}(\chi_{c2}\to \Sigma^{0}\bar{p}K^{+})}{\mathcal{B}(\chi_{c2}\to \Sigma^{+}\bar{p}K^{0}_{S})} = 1.10 \pm 0.13 \pm 0.07$, respectively, where common sources of  systematic uncertainties are canceled. These results are consistent with isospin symmetry within 1.6$\sigma$.

 Although there is no evident intermediate resonances on two-body subsystems of $\chi_{cJ}$ decays, the mass distributions of two-body subsystems are not completely consistent with the phase-space MC simulations. This implies the existence of intermediate baryon resonances. With the present statistics, it is difficult to study them in detail and draw any conclusions to them. More $\psi(3686)$ events in the future in combination with advanced analysis technique, such as partial wave analysis, may shed light on the intermediate structures.

\begin{acknowledgments}
The BESIII collaboration thanks the staff of BEPCII and the IHEP computing center for their strong support. This work is supported in part by National Natural Science Foundation of China (NSFC) under Contracts Nos. 11625523, 11635010, 11735014, 11822506, 11835012, 11935015, 11935016, 11935018, 11961141012, 11705006; the Chinese Academy of Sciences (CAS) Large-Scale Scientific Facility Program; Joint Large-Scale Scientific Facility Funds of the NSFC and CAS under Contracts Nos. U1732263, U1832207; CAS Key Research Program of Frontier Sciences under Contracts Nos. QYZDJ-SSW-SLH003, QYZDJ-SSW-SLH040; 100 Talents Program of CAS; INPAC and Shanghai Key Laboratory for Particle Physics and Cosmology; ERC under Contract No. 758462; German Research Foundation DFG under Contracts Nos. 443159800, Collaborative Research Center CRC 1044, FOR 2359, FOR 2359, GRK 214; Istituto Nazionale di Fisica Nucleare, Italy; Ministry of Development of Turkey under Contract No. DPT2006K-120470; National Science and Technology fund; Olle Engkvist Foundation under Contract No. 200-0605; STFC (United Kingdom); The Knut and Alice Wallenberg Foundation (Sweden) under Contract No. 2016.0157; The Royal Society, UK under Contracts Nos. DH140054, DH160214; The Swedish Research Council; U. S. Department of Energy under Contracts Nos. DE-FG02-05ER41374, DE-SC-0012069.

\end{acknowledgments}


\end{document}